\documentclass[
    ,final            % use final for the camera ready runs
  ]
  {aipproc}

\layoutstyle{8x11single}

\newcommand{\tmin}{\tau_{\rm min}}
\newcommand{\xmin}{x_{\rm min}}

\newcommand{\geqa}{\stackrel{>}{\scriptstyle \sim}}

\newcommand{\ef}{\epsilon_{\scriptscriptstyle F}}
\newcommand{\ec}{{\cal E}}

\newcommand{\ect}{\widetilde{\cal E}}

\newcommand{\rmb}{\rho_{\scriptscriptstyle MB}}
\newcommand{\rhob}{\overline{\rho}}
\newcommand{\deltab}{\overline{\delta}}
\newcommand{\rhot}{\widetilde{\rho}}

\newcommand{\scat}{\widetilde{{S}}}

\newcommand{\nc}{{\cal N}}

\newcommand{\taut}{\tau_{\scriptscriptstyle T}}
\newcommand{\xh}{x_{\rm H}}

\begin{document}

\title{Level density of a Fermion gas: average growth, fluctuations, universality}

\classification{21.10.Ma, 24.60.-k, 05.45.Mt \\ \\
Lecture delivered at the workshop ``Nuclei and Mesoscopic
Physics'', NSCL MSU, USA, October 23-26, 2004. To be published by American
Institute of Physics, V. Zelevinsky ed.}

\author{P. Leboeuf}{
  address={Laboratoire de Physique Th\'eorique et Mod\`eles Statistiques,
B\^at. 100, \\ Universit\'e de Paris-Sud, 91405 Orsay Cedex, France}}

\begin{abstract}
It has been shown by H. Bethe more than 70 years ago that the number of
excited states of a Fermi gas grows, at high excitation energies $Q$, like the
exponential of the square root of $Q$. This result takes into account only the
average density of single particle (SP) levels near the Fermi energy. It
ignores two important effects, namely the discreteness of the SP spectrum, and
its fluctuations. We show that the discreteness of the SP spectrum gives rise
to smooth finite--$Q$ corrections. Mathematically, these corrections are
associated to the problem of partitions of an integer. On top of the smooth
growth of the many--body density of states there are, generically,
oscillations. An explicit expression of these oscillations is given. Their
properties strongly depend on the regular or chaotic nature of the SP
motion. In particular, we analyze their typical size, temperature dependence
and probability distribution, with emphasis on their universal aspects.
\end{abstract}

\maketitle

%%%%%%%%%%%%%%%%%%%%%%%%%%%%%%%%%%%%%%%%%%%%
%% MAINMATTER
%%%%%%%%%%%%%%%%%%%%%%%%%%%%%%%%%%%%%%%%%%%%

\section{Introduction}

The main theoretical framework to understand the behavior of the density of
states of a Fermionic system has been, and still is, the independent particle
model. In the degenerate gas approximation, when the excitation energy of the
gas is much smaller than the Fermi energy $\ef$, the excitation spectrum
relies on the properties of the single--particle (SP) spectrum near $\ef$. In
most theoretical calculations only the average SP density of states, $\rhob$,
is taken into account. For a system of $A$ noninteracting fermions moving in a
mean--field potential, the number of excited states of the many--body (MB)
system contained in a small energy window $dQ$ at energy $Q$, $\rmb (A,Q) dQ$,
is \cite{bethe}
\begin{equation} \label{bethe}
\rmb (A,Q) = \frac{1}{\sqrt{48}\ Q} \exp \left( 2 \sqrt{a \ Q} \right) \ ,
\;\;\;\;\;\;\;\;\;\; a = \pi^2 \rhob/6 \ .
\end{equation}
Here $Q$ is measured with respect to the ground state energy of the gas. For a
given potential, $\rhob$ is in general a function of $A$. Besides the
condition $Q \ll \ef$, this expression assumes that the excitation energy is
large compared to the SP mean level spacing $\deltab = \rhob^{-1}$. For
simplicity, and because our aim here is not to compare with experimental data,
we ignore angular momentum conservation, isospin, etc.

After Bethe's work, more accurate calculations of $\rmb$ were made
\cite{bloch}. Schematic shell corrections related to a periodic fluctuation of
the SP density were computed in \cite{ros} (see also \cite{bm}), introducing
the so--called back-shifted Bethe formula. Several phenomenological
modifications of Eq.(\ref{bethe}) have been proposed to match the experimental
results. These models take into account, for example, shell effects, pairing
corrections and residual interactions \cite{gc,ist,krp,abf}, introducing a
multitude of coexisting phenomenological parameterizations. The present status
of the understanding do not allow to draw a clear theoretical picture of the
functional dependence of $\rmb$ with $A$ and $Q$.

Our purpose is, within a SP picture, to further develop the theoretical
analysis to include several important effects that are missing in
Eq.(\ref{bethe}). This presentation is based on the work reported in
\cite{lmr}.

There are different ways in which Eq.(\ref{bethe}) can be improved. As
mentioned before, it is the leading term of an expansion valid for a large
number of particles $A$ and for high energies $Q$ (compared to the SP mean
level spacing $\deltab$). For instance, assuming the gas is confined by a
Woods--Saxon like mean field potential, then
\begin{equation}\label{afermi}
\rhob = \frac{3}{2} \frac{A}{\ef} \; \Longrightarrow \; a = \frac{\pi^2}{4 \ef} A
\approx \frac{A}{15} \; {\rm MeV}^{-1} \ ,
\end{equation}
where we used the approximation $\ef \approx 37$ MeV. This expression of
$\rhob$ is the leading order term of an expansion in decreasing powers of $A$.
Corrections to it can be incorporated by considering lower order terms in the
Weyl series, that includes surface corrections, curvature corrections, etc
\cite{bh}. A similar expansion in decreasing powers of $A$ holds for the
energy or mass of the nucleus (the liquid drop formula). As is well known, the
coefficient of each term in the expansion of the energy obtained from a Fermi
gas is not correct, since the renormalization due to interactions is out of
the range of the model. Similarly, the coefficient $1/15$ in Eq.(\ref{afermi})
does not give a good description of the average trend of the experimental data
obtained from neutron resonances, which is closer to $1/8$ (cf Fig.1).
Although of great interest, we will not discuss these $A$ corrections and
their renormalization due to interactions, but concentrate on two other
corrections that are within the scope of a noninteracting model. The first one
(next section) leads to smooth lower order terms in the excitation energy $Q$.
The second type of corrections incorporates oscillatory terms in $A$ and $Q$.
These oscillations, superimposed to the smooth $A/8$ trend mentioned above,
are clearly visible in the experimental data shown in Fig.1.

\begin{figure}
  \includegraphics[height=.36\textheight,width=.52\textheight]{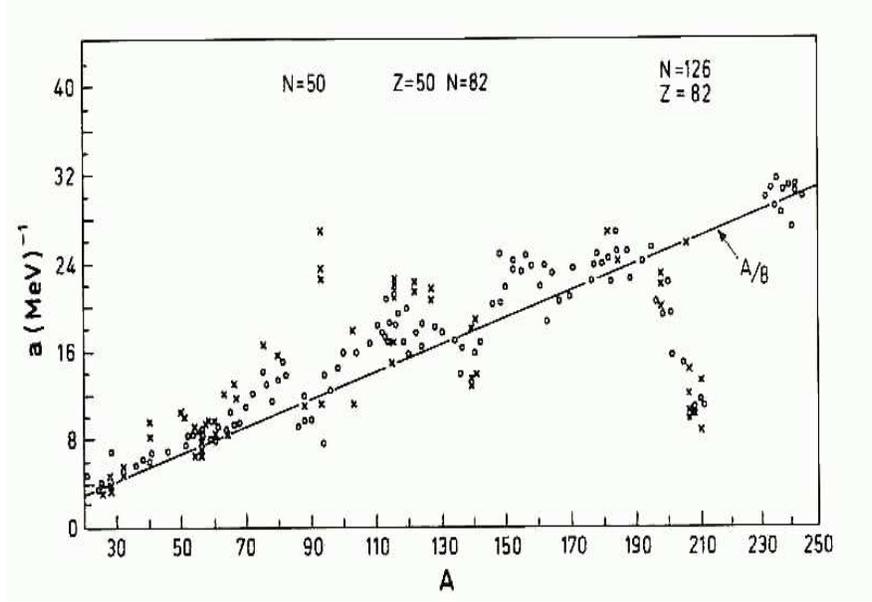}
  \caption{The parameter $a$ appearing in the Fermi gas level density formula
  as a function of the mass number $A$ (taken from Ref.\cite{bm}).}
\end{figure}

\section{Discreteness of the spectrum: average behavior}

The only property of the SP spectrum in Eq.(\ref{bethe}) is the average
density of levels $\rhob$. This approximation treats the SP spectrum as a
continuum, and ignores the influence in the many--body density of states of
the discreteness of the spectrum and of the exact position of the SP levels.
One step beyond, that incorporates the discreteness but ignores the
fluctuations, is to consider a locally perfectly regular SP spectrum (1D
harmonic oscillator), given by the SP energies $\epsilon_n = n \delta$, where
$n$ is an arbitrary integer, and $\delta$ is the distance between neighboring
levels. The Fermi gas model consists of $A$ noninteracting fermions that
occupy the equidistant levels, with occupation number 0 or 1. In the ground
state the particles fill the lowest $A$ SP states. We restrict the analysis to
the approximation $Q \ll \ef$, in which the number of excited particles is
small compared to the total number $A$, and effects due to a finite number of
particles can be ignored. In the excited states particles occupy SP levels
above $\ef$, thus creating holes in the Fermi sea. The possible excited
energies, measured with respect to the ground state energy, are $Q = m
\delta$, where $m$ is an arbitrary positive integer. The excitation energies
are thus trivial. The nontrivial information comes from the degeneracy of each
of theses energies, since there are many different many--body configurations
with the same excitation energy. The problem then is how to compute the
degeneracy of each excited state of energy $Q= m \delta$.

\begin{figure}
  \includegraphics[height=.24\textheight]{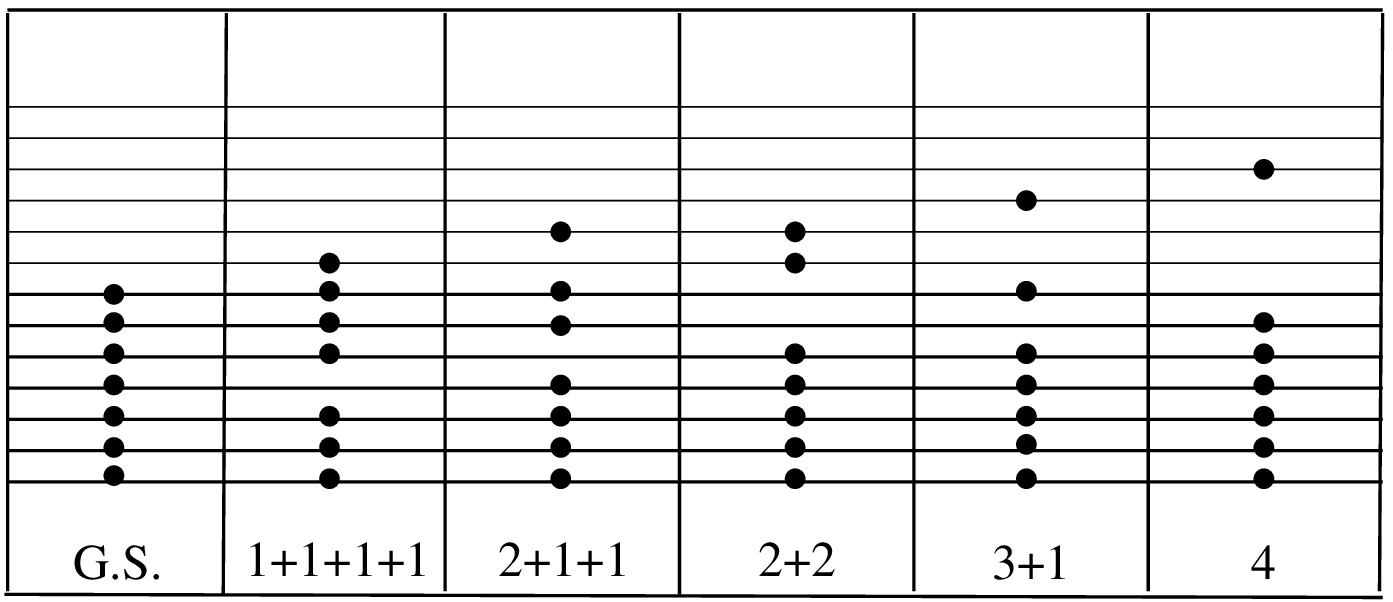}
  \caption{Ground state (G.S., left column) and excited states of energy
  $Q/\delta = 4$ (next five columns) of a gas of noninteracting fermions in a
  perfectly regular SP spectrum. Each excited state is identify with a
  decomposition of the excitation energy 4 as a sum of integers (bottom of each
  column).}
\end{figure}

The solution is obtained by realizing that the problem we are facing is
exactly equivalent to a well known problem in mathematics, i.e. in how many
different ways an integer can be decomposed as a sum of integers (partitions
of an integer, $p(m)$). We will simply provide here a graphical illustration
of this equivalence (Fig.2). Consider for example the case $m=4$, the argument
easily generalizes to any $m$. $m=4$ can be decomposed as a sum of integers in
5 different ways (and thus $p(4)=5$): $1+1+1+1$, $2+1+1$, $2+2$, $3+1$ and
$4$. As Fig.2 shows, each of these decompositions is in a one-to-one
correspondence with one of the possible excited states of the Fermi gas of
total energy $Q/\delta = 4 $. The ground state is shown on the left column. In
the second column from the left four particles have been pushed one level up,
hence the equivalence with $1+1+1+1$ (other mappings may be used as well). In
the third column the upper particle was pushed up by two, the next two by one.
Etc. As has been emphasized many times, there is also a (trivial) one--to--one
correspondence between the partitions of an integer and the excited states of
a gas of noninteracting bosons in a one--dimensional harmonic oscillator
potential. In this case each integer in the decomposition of the excitation
energy indicates the level occupied by one of the excited particles.

The partition of an integer grows very fast with $m$: it is 5 for $m=4$,
whereas $p(200) = 397999029388$ (this number was computed by hand by P.
MacMahon in 1918). The generating function of $p(m)$ was obtained in 1753 by
L. Euler. An explicit expression for $p(m)$ came much later, under the form of
an asymptotic (exact) formula (initially obtained by Hardy and Ramanujan
\cite{hr}, improved and made rigorous later on by Rademacher
\cite{rademacher}). To facilitate the comparison with more general approaches,
we express the result in terms of $\rhob Q$, where here $\rhob = \delta^{-1}$
is the inverse of the exact spacing between neighboring levels. The first
terms of the expansion are given by,
\begin{equation} \label{shr2}
\log \left( \rmb^{\scriptscriptstyle HO}/\rhob \right) = \sqrt{ \frac{2}{3}
\pi^2 \rhob \ Q } - \log \left( \sqrt{48} \ \rhob Q \right) -  \
\frac{\pi^2 + 72}{24 \sqrt{6} \pi} (\rhob Q)^{-\frac{1}{2}} - \left(\frac{3}{4
\pi^2}- \frac{1}{24} \right) (\rhob Q)^{-1} + {\cal O} ((\rhob
Q)^{-\frac{3}{2}}) \ .
\end{equation}
We emphasize with the notation $\rmb^{\scriptscriptstyle HO}$ that this is the
many--body density of states for a Fermi gas in a 1D harmonic oscillator
spectrum. As expected, the first two terms of this equation reproduce Bethe's
formula. However, Eq.(\ref{shr2}) goes much further: it is a truncation of the
{\sl exact} asymptotic expansion of the density of states, whose precision can
be improved by adding further terms. Notice that all the terms in
Eq.(\ref{shr2}) are smooth functions of the excitation energy $Q$.

It is natural to ask about the relevance of Eq.(\ref{shr2}) in realistic
systems, where the true SP spectrum consists of discrete energy levels
arranged, in most cases, with no particular order. Equation (\ref{shr2})
clearly goes beyond Bethe's result, since it describes the exact behavior of
the many--body density of states of a gas of particles that occupy a discrete,
perfectly regular arrangement of SP levels. It is therefore not unreasonable
to expect that the result is valid for an arbitrary spectrum having the same
average density $\rhob$, and that the effects of the fluctuations of the SP
energy levels with respect to a perfectly regular arrangement come on top of
it. Although we do not have for the moment an explicit proof of this
statement, numerical simulations to be presented below seem to confirm this
hypothesis.

\section{Fluctuations}

The exact dependence of $\rmb$ on $A$ and $Q$ is sensitive to the detailed
arrangement of the SP energy levels around the Fermi energy. Strong deviations
with respect to a regular spacing, with possible degeneracies, may be produced
by the presence of symmetries of the confining self--consistent potential.
These deviations induce, in turn, oscillations in the thermodynamic functions
of the gas. Shell effects are therefore due to deviations (or bunching) of the
SP levels with respect to a perfectly regular spectrum. The degeneracies of
the electronic levels of an atom produced by the rotational symmetry are an
extreme manifestation of level bunching. In general, in systems that have
other symmetries, or have no symmetries at all, there will still be level
bunching, but its importance will typically be minor. Therefore, depending on
the presence or absence of symmetries, the shell effects may be more or less
important. The level bunching, and more generally the fluctuations of the SP
energy levels, are thus a very general phenomenon. The theories that describe
those fluctuations make a neat distinction between systems with different
underlying classical dynamics (e.g., regular or chaotic). Our understanding of
these fluctuations and of their connections with the regular or chaotic nature
of the SP motion has greatly increased in the last decades. For a recent
review see \cite{sev}.

Oscillations in the many--body density of states of nuclei are clearly visible
in the experimental data available from neutron resonances as a function of
the mass number $A$. The parameter $a$ extracted from the data, plotted in
Fig.1, shows oscillations around an average growth ($A/8$). The effective
description of the oscillations through the parameter $a$ is an artifact of
the analysis, since the Fermi gas model relates $a$ to $\rhob$, a smooth
function. As we will see below, the theoretical analysis leads to oscillatory
corrections that enter as an additional term in the exponent of $\rmb$, and
thus cannot in general be interpreted as an effective $a$.

There are two distinct and important scales to describe the SP fluctuations.
The first one, the smallest one, is the mean spacing between SP energy levels
$\deltab$. The fluctuations of the SP energy levels on that scale have been
shown to be, for a large class of systems, {\sl universal}. Their statistical
properties are described by uncorrelated sequences for integrable systems, and
by random matrix theory in chaotic ones. The second relevant energy scale is
$E_c = h/\tmin$, where $\tmin$ is of the order of the time of flight across
the system (a precise definition will be given below). For $A$ large, the
ratio $g = E_c /\deltab$ is much larger than one (for instance, $g\sim
A^{2/3}$ in a three dimensional cavity). In contrast to the fluctuations on
scales $\deltab$, the fluctuations on scales $E_c$ are long range modulations
of the SP spectrum whose structure and amplitude are system specific. There is
therefore no universality on this scale.

For an arbitrary SP spectrum the computation of the density of excited states
of a fermionic system is a difficult combinatorial problem for which no exact
solution is known. What we are seeking here are the variations of the MB
density of states due to fluctuations of the SP spectrum with respect to a
perfectly ordered spectrum, described in the previous section. An explicit
formula that includes these effects has been obtained recently \cite{lmr}. The
result, obtained from a saddle point approximation of an inverse Laplace
transform of the MB density, takes the form \cite{lmr}
\begin{equation} \label{rhodef}
\rmb (A,Q) = \frac{1}{\sqrt{48}\ Q} \exp \left\{ 2 \sqrt{a \ Q}
+ \sqrt{\frac{a}{Q}} \left[ \ect (\mu,0) - \ect (\mu,T) \right] \right\} \ .
\end{equation}
The function $\ect (\mu,T)$ is the fluctuating part of the energy of
the gas at chemical potential $\mu$ and temperature $T$, defined as follows.
The energy is, as usual, given by
\begin{equation} \label{u}
\ec (\mu,T) = \int d\epsilon \ \epsilon \ \rho (\epsilon) \left[ 1 + {\rm
        e}^{(\epsilon-\mu)/T} \right]^{-1} \ ,
\end{equation}
where the function $\rho (\epsilon)$ is the SP density of states defined by
the SP energies $\epsilon_i$
$$
\rho (\epsilon) = \sum_i \delta (\epsilon-\epsilon_i) \ .
$$
The SP density of states is usually decomposed into a part $\rhob$ that has a
smooth dependence on the energy $\epsilon$ plus another contribution $\rhot$
that describes the deviations with respect to the smooth behavior. The
fluctuating part of the energy is defined as
\begin{equation}\label{ect}
\ect (\mu,T) = \ec (\mu,T) - \int d\epsilon \ \epsilon \ \rhob (\epsilon)
        \left[ 1 + {\rm e}^{(\epsilon-\mu)/T} \right]^{-1} \ .
\end{equation}

In Eq.~(\ref{rhodef}) the arguments of $\ect$ are functions of $A$ and $Q$,
$\mu = \mu(A,Q)$ and $T=T(A,Q)$. These two functions are determined by
inversion of the equations
 \begin{eqnarray}\label{spc}
A &=& \nc (\mu, T) \ , \\
Q &=& \ec (\mu, T) - \ec (\mu,0) \ . \nonumber
\end{eqnarray}
The function $\nc (\mu,T) = \int d\epsilon \ \rho (\epsilon) \left[ 1 + {\rm
e}^{(\epsilon-\mu)/T} \right]^{-1}$ is the particle number. We are ignoring
here (small) variations of the chemical potential with temperature. To lowest
order, in which only the average behavior of the energy and of the particle
number is taken into account, the second equation in (\ref{spc}) leads to the
usual relation between temperature and excitation energy
\begin{equation}\label{tq}
T \approx \sqrt{\frac{Q}{a}} \ ,
\end{equation}
and $\mu$ is simply a function of $A$, $\mu=\mu(A)$. For instance, for a Fermi
gas in a 3D cavity of volume $V$ and particle mass $m$,
\begin{equation}\label{mua}
\mu \approx \frac{2 \pi \hbar^2}{m} \left( \frac{3 \sqrt{\pi} A}{4 V} \right)^{2/3} \ .
\end{equation}

Equation (\ref{rhodef}) shows that, superimposed to the smooth growth of the
density described by Eq.(\ref{bethe}), there are oscillatory corrections in
the exponent of the density of states. These corrections are directly related
to energy fluctuations of the system.

The oscillatory nature of the new corrections is clearly displayed through a
semiclassical theory, that expresses the quantum properties of the fermion gas
in terms of classical solutions of the SP equations of motion. In this
approach $\ect (\mu,T)$ is written as a sum over the classical periodic orbits
of the mean--field potential \cite{sm,lm3,sev}
\begin{equation} \label{grandosc}
\ect (\mu,T)= 2 \hbar^2 \sum_p \sum_{r=1}^{\infty} \frac{A_{p,r}
~ \kappa ( r ~ \tau_p/\taut)}{r^2 ~ \tau_p^2} \cos \left(
r \ S_p /\hbar + \nu_{p,r} \right) \ .
\end{equation}
Each orbit is characterized by its action $S_p$, stability amplitude
$A_{p,r}$, and Maslov index $\nu_{p,r}$. $\tau_p$ is the period of the
periodic orbit, and $\kappa (x) = x/\sinh (x) $ is a temperature factor that
introduces the time scale $\taut = \hbar/(\pi T)$ conjugate to the
temperature. This expression describes the departures of $\ec$ with respect to
its mean behavior due to the fluctuations of the SP spectrum. The behavior of
$\ect$ strongly depends on whether $\mu$ or $T$ (or $A$ or $Q$) are varied. A
temperature variation modifies the prefactors of the summands in $\ect$
(through the function $\kappa$), and therefore produces gentle variations of
the fluctuating part in the exponent of the density of states
\begin{equation}\label{scat}
\scat (A,Q) = \sqrt{\frac{a}{Q}} \left[ \ect (\mu,0) - \ect (\mu,T) \right] \ .
\end{equation}
In contrast, $A_{p,r}$, $\tau_p$ and $S_p$ depend on $\mu$ (and therefore on
$A$). For large values of $\mu$, $S_p \gg \hbar$ and the dominant variation
with the particle number (or any other parameter that modifies the actions)
comes from the argument of the cosine function in $\ect$. Rapid oscillations
of $\scat$ are therefore generically expected when the number of particles is
varied.

\begin{figure}
\includegraphics[width=8.5cm,height=7.0cm]{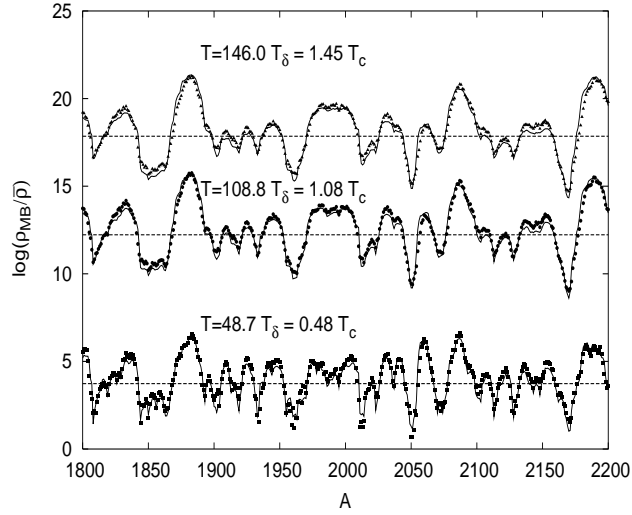}
\caption{The logarithm of the many-body density of states of $A$
     noninteracting fermions in a rectangular billiard of sides $a = \sqrt{(1
     + \sqrt{5})/2}$ and $b=a^{-1}$. The value of $g$ is 100. Dots: numerical
     computation at three different excitation energies; solid curves:
     theoretical prediction (\ref{hro}); dashed curves: smooth part
     Eq.(\ref{shr2}).}
\end{figure}

For a given mean field potential there exists an infinite number of periodic
orbits $p$. The spectrum of periods $\tau_p$ has no upper bound. It has,
however, a lower bound, given by the period $\tmin$ of the shortest periodic
orbit. Since $\tmin$ is usually the smallest characteristic time scale in the
system, it determines the largest energy scale $E_c = h/\tmin$ in which
modulations (bunching) of the SP energy levels occur

If the classical periodic orbits are known in a particular system, the
corrections can be computed explicitly. The periodic--orbit sum is dominated
by the short orbits, and rapidly gives a good approximation of the result. To
illustrate our findings we have tested some of the predictions by a direct
numerical counting of the MB density of states in a particular system. Figure
3 shows the results obtained for a gas of about 2000 fermions contained in a
two--dimensional rectangular cavity, an integrable system. For each number of
particles we compute the MB density of states at three different temperatures,
measured in units of
$$
T_{\delta}=\deltab/2 \pi^2 \;\; {\rm and} \;\; T_c =E_c /2 \pi^2 \ .
$$
The theoretical curve is computed according to the expression
\begin{equation} \label{hro}
\log ( \rmb/\rhob ) = \log ( \rmb^{\scriptscriptstyle HO}/\rhob ) + \scat
(A, Q) \ ,
\end{equation}
where $\scat$ is calculated from the periodic orbits of the rectangle.

If the smooth corrections to the density of Eq.(\ref{shr2}) are not included,
and only the result of Bethe is used, a systematic deviation is observed in
the average behavior of $\rmb$ between theory and numerics. Although the
theoretical analysis of Ref.\cite{lmr} does not include those corrections, the
numerical results seem to indicate that their validity goes beyond the simple
1D harmonic oscillator spectrum. Notice the extremely good accuracy of
Eq.(\ref{hro}), either for the average value of the density as well as for the
fluctuations. Notice also that the typical size of the fluctuations can be
quite important (it is of the order of the average for the lowest excitation
energy computed).

\section{Statistical analysis of the fluctuations: universality}

Equation (\ref{hro}) decomposes the exponent of the density of states into a
smooth part (given by the Hardy--Ramanujan result) plus oscillatory
contribution, $\scat (A,Q)$, associated in a semiclassical theory to a sum
over periodic orbits. $\scat (A,Q)$ relates the shell corrections in the MB
density of states to the fluctuations of the energy of the gas. $\ect$ depends
on temperature only through the function $\kappa$ (cf Eq.(\ref{grandosc})).
The effect of this function is to exponentially suppress the contribution of
periodic orbits whose period $\tau_p \gg \taut$ \cite{ruj,lm3}. Since there is
no suppression at $T=0$ (because $\taut \rightarrow \infty$), only orbits
whose period $\tau_p \geqa \taut$ contribute to the difference $\ect (\mu,0) -
\ect (\mu,T)$. At temperatures such that $\taut \ll \tmin$, the term $\ect
(\mu,T)$ becomes exponentially small, and only $\ect (\mu,0)$ remains. The
shell correction $\scat (A,Q)$ therefore decays as $T^{-1}$ at $T \gg T_c$.
This decay contrasts with the more common exponential damping observed in
other thermodynamic quantities \cite{lm3}.

A clearer picture of how the fluctuations behave may be obtained through a
statistical analysis. The most simple property of the fluctuations is $\langle
\scat \rangle =0$, where the brackets denote an average over a suitable
chemical potential (or particle number) window. This result is valid only to
first order in the expansion leading to Eq.(\ref{rhodef}) \cite{lmr}; it can be
shown that higher order terms contribute to a non--zero average. The next
non--trivial statistical property is the variance $\langle \scat^2 \rangle$,
that may be computed using Eq.(\ref{grandosc}). The result is $\langle \scat^2
\rangle = (1/2) \int_{0}^\infty d x \ K(x,\xh) \left[ 1 - \kappa (x) \right]^2
/x^4$, where $K(x,\xh)$ is the rescaled form factor of the SP spectrum (cf
Eq.(36) in Ref.\cite{lm3}). The latter function depends on the rescaled
Heisenberg time $\xh=h \rhob/\taut$. It describes system--dependent features
for $x$ of the order of $\xmin = \tmin/\taut$, while it is believed to be
universal for $x \gg \xmin$. The universality class depends on the regular or
chaotic nature of the dynamics, and on its symmetry properties. 

Taking into account the basic properties of $K(x,\xh)$, in {\bf chaotic
systems} three different regimes for $\langle \scat^2 \rangle$ as a function
of temperature are found \cite{lmr} (remember the relation (\ref{tq}) between
$T$ and $Q$):

\begin{itemize}

\item Low temperatures $T \ll T_\delta$. In this regime 
\begin{equation} \label{ltr}
\langle \scat^2 \rangle = c_4 T/(2 T_\delta) \ ,
\end{equation}
where $c_4 = 0.0609...$ is the value at $k=4$ of the constant
\begin{equation}\label{ck}
c_k = \int_0^\infty \frac{d \ x}{x^k} \left[ 1 - \frac{x}{\sinh x} \right]^2 \ .
\end{equation}

\item Intermediate temperatures $T_\delta \ll T \ll T_c$. In
this regime the size of the fluctuations saturates at a universal
constant 
\begin{equation} \label{itr}
\langle \scat^2 \rangle = c_3/\beta \ ,
\end{equation}
where $c_3 = 0.1023\ldots$ and $\beta = 1 (2)$ for systems with (without)
time--reversal invariance.

\item High temperatures $T \gg T_c$. The size of the fluctuations decreases
with excitation energy, 
\begin{equation} \label{htr}
\langle \scat^2 \rangle = \langle \ect^2 (\mu,0) \rangle [ 1 - 8 \ {\rm
e}^{-T/T_c} ]/ T^2 \ .
\end{equation}
After an exponential transient, a power--law decay $\langle \scat^2
\rangle^{1/2} \propto T^{-1}$ is thus obtained.

\end{itemize}

The situation is different in {\bf integrable systems}, where only two regimes
are found. At low temperatures the result is identical to that of chaotic
systems. The linear growth at low temperatures is thus totally universal and
independent of the system. The difference is that in integrable systems the
growth extends up to much higher temperatures, $T \approx T_c$, without
saturation. At that temperature the variance of the fluctuations is of order
$g$. In integrable systems, the maximum amplitude of the fluctuations is
therefore reached at $T \approx T_c$, and its typical size is much larger than
in chaotic systems. At high temperatures $T \gg T_c$ the decay is almost
identical to that of chaotic systems (the coefficient 8 is replaced by a 12).
A schematic representation of the temperature dependence of the variance of
the fluctuations for chaotic and integrable systems is given in Fig.4.

\begin{figure}
  \includegraphics[height=.24\textheight,width=.54\textheight]{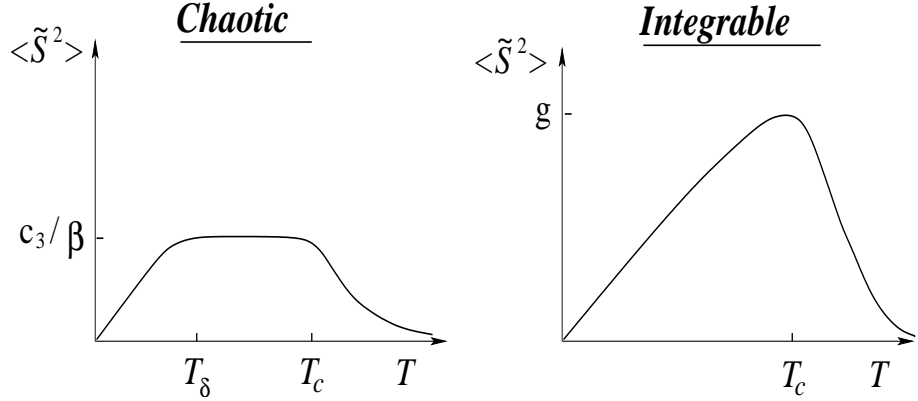}
  \caption{Schematic representation of the temperature dependence of the
  variance of the fluctuations in chaotic and integrable systems. In each
  class, the behavior is universal in the regime $T \ll T_c$.}
\end{figure}

It has been shown \cite{lm3} that $\ect (\mu,T)$ is dominated, at any $T$, by
the shortest classical periodic orbits. In contrast, the difference $\ect
(\mu,0)-\ect (\mu,T)$ depends on orbits whose period $\tau_p \geqa \taut$. For
temperatures $T \ll T_c$ the statistical properties of these orbits are
universal, and correspondingly the probability distribution function of
$\scat$ is expected to be universal, in the sense that at a given temperature
it should only depend on the nature of the underlying classical dynamics
(regular or chaotic), and the symmetries of the system. This statement is
supported by the fact that in the limit $T \rightarrow 0$, $\scat \approx -
\partial \ect (\mu,T) /\partial T$. The probability distribution of the latter
quantity was studied in Ref.\cite{lm3}; it was shown that it coincides at low
temperatures with that obtained from a Poisson spectrum for integrable systems
and from a random matrix spectrum for chaotic ones. This confirms the
universality of the distribution at low temperatures. As the temperature is
raised, the universality of the probability distribution of $\scat$ will be
lost for temperatures of the order or greater than $E_c$, where system
specific features (i.e., short periodic orbits) are revealed. In this respect,
the decay (\ref{htr}) is only indicative. Its exact form depends on the
details of the short periodic orbits spectrum, treated here only through a
rough approximation.

\section{Concluding remarks}

Two improvements with respect to Bethe's formula of $\rmb$ have been
discussed. The first one incorporates the discreteness of the SP spectrum by
considering a set of equidistant SP states (1D harmonic oscillator). The exact
many--body density of states is obtained by mapping the problem to the
computation of the number of decompositions of an integer as a sum of
integers. The exact formula adds finite--$Q$ corrections to the asymptotic
result. Although a proof is missing, numerical simulations suggest that the
range of validity of this result extends beyond the 1D harmonic oscillator
spectrum to any spectrum, in the sense that it describes with good accuracy
the smooth part of the growth of $\rmb$ in systems with an arbitrary SP
spectrum (cf Fig.3).

Beyond the smooth behavior, the second improvement concerns the fluctuations
or shell effects in $\rmb$. In a semiclassical theory, each periodic orbit has
been shown to contribute to $\log (\rmb/\rhob)$ with a fluctuating term as a
function of the number of particles $A$, of wavelength
$$
\Delta A = \frac{h/\tau_p}{\deltab} \ .
$$ 
At low excitation energies long orbits contribute to the fluctuations while
short ones are exponentially suppressed, leading to wild oscillations and
universality of their statistical properties. As the temperature increases,
the situation is reversed, long orbits are exponentially suppressed while
short orbits come into play. At $T=T_c$, oscillations of wavelength of order
$\Delta A = E_c/\deltab = g$ are predicted. These oscillations are clearly
visible, with the correct wavelength, in Fig.3. With the raise of the short
periodic orbits as $T$ increases, the universality of the statistical
properties of the fluctuations disappears for $T$ of the order or higher than
$T_c$.

Concerning the typical size of the oscillations, a nontrivial dependence as a
function of excitation energy (or temperature) was found. The results are
summarized in Fig.4. The behavior is quite different in chaotic and integrable
systems. At low temperatures the variance grows linearly with $T$ in all
cases. However, a plateau is rapidly reached at $T=T_\delta$ in chaotic
systems, whereas the amplitude of the fluctuations continue to grow in regular
ones. At $T=T_c$ the differences reaches its maximum amplitude, where the
variance in regular systems is $g$ times larger than in chaotic ones
(remember, moreover, that $g\sim A^{2/3}$ in a three dimensional cavity, so
that the difference increases with an increasing number of particles). At this
point the situation is similar to what was found for the ratio of the variance
of the fluctuations of the nuclear mass due to regular and chaotic motion
\cite{bl}. At $T \gg T_c$ the variance decreases as $T^{-2}$ for both types of
dynamics. The validity of the results discussed here in the analysis and
interpretation of the experimental data on the nuclear level density will be
presented elsewhere.

%%%%%%%%%%%%%%%%%%%%%%%%%%%%%%%%%%%%%%%%%%%%%%%%
%% BACKMATTER
%%%%%%%%%%%%%%%%%%%%%%%%%%%%%%%%%%%%%%%%%%%%%%%%

\begin{theacknowledgments}
I thank Professor Vladimir Zelevinsky and Ms Shari Conroy for their kind
hospitality at the workshop ``Nuclei and Mesoscopic Physics'' in East Lansing,
October 2004.
\end{theacknowledgments}

%%%%%%%%%%%%%%%%%%%%%%%%%%%%%%%%%%%%%%%%%%%
%% The following lines show an example how to produce a bibliography
%% without the help of the BibTeX program. This could be used instead
%% of the above.
%%%%%%%%%%%%%%%%%%%%%%%%%%%%%%%%%%%%%%%%%%%

\end{document}